\documentclass[a4paper,12pt]{article}

\newcommand{\sect}[1]{\setcounter{equation}{0}\section{#1}}

\begin{document}
\topmargin 0pt \oddsidemargin 0mm

\renewcommand{\thefootnote}{\fnsymbol{footnote}}
\begin{titlepage}
\begin{flushright}
\end{flushright}

\vspace{1mm}
\begin{center}
{\Large \bf Ricci Flat Black Holes and Hawking-Page Phase Transition
in Gauss-Bonnet Gravity and Dilaton Gravity }

 \vspace{10mm}
{\large
Rong-Gen Cai\footnote{Email address: cairg@itp.ac.cn}}\\
\vspace{5mm}
{ \em Institute of Theoretical Physics, Chinese Academy of Sciences, \\
   P.O. Box 2735, Beijing 100080, China}\\
\vspace*{.5cm}
 {\large Sang Pyo Kim\footnote{Email address:
 sangkim@kunsan.ac.kr}}\\
 \vspace*{5mm}
 {\em Department of Physics, Kunsan National University, Kunsan 573-701,
 Korea}\\
{\em Asia Pacific Center for Theoretical Physics, Pohang 790-784,
Korea} \vspace*{0.5cm} \\
 {\large Bin Wang~\footnote{Email address: wangb@fudan.edu.cn}}\\
 \vspace*{5mm}
 {\em Department of Physics, Fudan University, Shanghai 200433,
 China}

\end{center}
\vspace{5mm} \centerline{{\bf{Abstract}}}
 \vspace{5mm}
It is well-known that there exists a Hawking-Page phase transition
between a spherical AdS black hole and a thermal AdS space.  The
phase transition does not happen between a Ricci flat AdS black hole
whose horizon is a Ricci flat space and a thermal AdS space in the
Poincare coordinates. However, the Hawking-Page phase transition
occurs between a Ricci flat AdS black hole and an AdS soliton if at
least one of horizon coordinates for the Ricci flat black hole is
compact. We show a similar phase transition betwen the Ricci flat
black holes and deformed AdS solitons in the Gauss-Bonnet gravity
and the dilaton gravity with a Liouville-type potential including
the gauged supergravity coming from the spherical reduction of
Dp-branes in type II supergravity. In contrast to Einstein gravity,
we find that the high temperature phase can be dominated either by
black holes or deformed AdS solitons depending on parameters.

\end{titlepage}

\newpage
\renewcommand{\thefootnote}{\arabic{footnote}}
\setcounter{footnote}{0} \setcounter{page}{2}
\sect{Introduction}

 Since the AdS/CFT  correspondence was proposed~\cite{AdS}, a lot
 of attention has been focused on
the black holes in AdS space, and various properties of black holes
in AdS space have been studied. In the spirit of the AdS/CFT
correspondence, Witten~\cite{Witten} has argued that the
thermodynamics of black holes in AdS space can be identified with
that of dual strong coupling CFTs in high temperature limit.
Therefore one can discuss the thermodynamics and phase structure of
strong coupling CFTs by studying the thermodynamics of various kinds
of black holes in AdS space. Indeed, it is well-known that there
exists a phase transition between the Schwarzschild-AdS black hole
and thermal AdS space, the so-called Hawking-Page phase
transition~\cite{HP}: the black hole phase dominates the partition
function in a high temperature limit, while the thermal AdS space
dominates in a low temperature limit.  That is, in the AdS space,
thermal gas will collapse to form a stable black hole when
temperature increases. This phase transition is a first order one,
and is interpreted as the confinement/deconfinement phase transition
in the dual CFTs~\cite{Witten}.

One of remarkable properties of black holes in AdS space is that
the black hole horizon is not necessarily a sphere~\cite{Topo}. The
case of black hole horizon being a Ricci flat surface was first
discussed in \cite{Flat}. The black hole horizon can also be a
negative constant curvature surface~\cite{Negative}. These
so-called topological black holes have been investigated in higher
dimensions \cite{Birm,high,GB,Love} and in dilaton
gravity~\cite{CJS,CZ}. It was found that the Hawking-Page phase
transition, which happens for spherical AdS black holes, does not
occur for Ricci flat and negative curvature AdS black holes, the
latter two being not only locally stable (heat capacity is always
positive), but also globally stable (see for example,
\cite{Birm}). Note that to see whether a black hole is globally
stable and a phase transition happens, one has to calculate
the Euclidean action of the black hole. As is well-known, the
gravitational action  always diverges due to an infinite space. To
get a finite result, one usually takes two different approaches:
one is the surface counterterm method, in which some surface terms
are added to result in a finite action; the other is called
the background subtraction method in which a suitable reference background is
chosen so that the solution under study can be asymptotically embedded into
the reference background.

For the Ricci flat AdS black hole, which is the main topic of this
paper, the conclusion that the solution is globally stable and no
Hawking-Page phase transition happens, is drawn by choosing the AdS
space in the Poincare coordinates as a reference background (the
surface counterterm method is equivalent to choosing the reference
background in this case). That is, one views the AdS space in the
Poincare coordinates, a reference vacuum, as the lowest energy
state. Remarkably, as Horowitz and Myers~\cite{HM} showed that there
does exist another kind of gravitational configuration, which has
lower energy than the AdS space in the Poincare coordinates, but
with the same boundary topology as the Ricci flat black hole and the
AdS space in the Poincare coordinates. The new configuration is
called the AdS soliton. Regarding the AdS soliton as a reference
background, Surya, Schleich and Witt~\cite{SSW} found that a phase
transition will happen between the Ricci flat AdS black hole and the
thermal AdS soliton if at least one of black hole horizon
coordinates is compact (see also \cite{Page,Myers}). The latter is
required in order that the black hole can be asymptotically embedded
into the background. The compact direction of the horizon plays a
crucial role in the Hawking-Page phase transition. The possibility
of quasi-normal modes as a probe to the Hawking-Page phase
transition has been discussed more recently~\cite{Shen}.

In this paper, we discuss the Hawking-Page
phase transition between Ricci flat black holes and deformed AdS
soliton in the Gauss-Bonnet gravity and dilaton gravity with a
Louville-type potential including the gauged supergravity comings
from the spherical reduction of Dp-branes in type II supergravity.
The effect of Gauss-Bonnet term on the Hawking-Page phase
transition of spherical AdS black holes has been studied in
\cite{GB}. In the next section, we first discuss the Ricci flat
black holes and deformed AdS soliton in the Gauss-Bonnet gravity,
and see the effect of the Gauss-Bonnet term on the phase
transition between the Ricci flat black hole and thermal AdS
soliton. In Sec.~3, we consider the Ricci flat black holes and
associated deformed AdS soliton in dilaton gravity. The conclusion
is given in Sec.~4.


\sect{Deformed AdS Soliton and Ricci flat black holes in Lovelock
gravity}

 In this section we study Ricci flat black hole and AdS soliton in
  Gauss-Bonnet gravity in $p+2$-dimensions, whose action
 is given by
 \begin{equation}
 \label{2eq1}
 S= \frac{1}{16\pi G} \int d^{p+2}x \sqrt{-g} \left( R +\frac{p
  (p+1) }{l^2} +\alpha R_{GB}\right),
  \end{equation}
  where $R_{GB}= R_{\alpha\beta \gamma\delta}R^{\alpha\beta
  \gamma\delta} -4R_{\mu\nu}R^{\mu\nu} +R^2$ is the Gauss-Bonnet
  term,  $\alpha$ is called the Gauss-Bonnet coefficient with
  dimension $(length)^2$, and $l^{-2}$ is related to the
  cosmological constant $\Lambda= -p(p+1)/2l^2$. Varying the action
  yields the equations of motion
  \begin{eqnarray}
  \label{2eq2}
 R_{\mu\nu}-\frac{1}{2}g_{\mu\nu}R &= &\frac{p(p+1)}{2l^2}g_{\mu\nu}
  + \alpha \left (\frac{1}{2}g_{\mu\nu}(R_{\gamma\delta\lambda\sigma}
  R^{\gamma\delta\lambda\sigma}-4 R_{\gamma\delta}R^{\gamma\delta}
  +R^2) \right. \nonumber \\
 &&~~~- \left. 2 RR_{\mu\nu}+4 R_{\mu\gamma}R^{\gamma}_{\ \nu}
  +4 R_{\gamma\delta}R^{\gamma\  \delta}_{\ \mu\ \ \nu}
   -2R_{\mu\gamma\delta\lambda}R_{\nu}^{\ \gamma\delta\lambda} \right).
\end{eqnarray}
The Ricci flat black hole in the Gauss-Bonnet gravity is~\cite{GB}
\begin{equation}
\label{2eq3}
 ds^2=-V_b(r) dt^2 +V_b(r)^{-1} dr^2 +r^2 (dx^2 +h_{ij}dx^idx^j),
 \end{equation}
 where~\footnote{In fact, there are two branches for the solution. One branch is
 argued to be unstable, so we will not discuss that
 branch~\cite{GB}.}
 \begin{equation}
 \label{2eq4}
 V_b(r) =\frac{r^2}{2\tilde\alpha}\left ( 1 -
 \sqrt{1+\frac{64 \pi G\tilde \alpha M_b}{p\eta_b \Sigma r^{p+1}}-
   \frac{4\tilde\alpha}{l^2}} \right),
   \end{equation}
$\tilde \alpha= (p-1)(p-2)\alpha$, $M_b$ is an integration constant
and is related to the mass of the black hole, $\eta_b$ is the period
of the coordinate $x$ (suppose it is compact), and $\Sigma$ is the
volume of the $(p-1)$-dimensional Ricci flat space described by
$h_{ij}dx^idx^j$ with topology ${\cal M}_{p-1}= R^{p-1}/\Gamma$,
where $\Gamma$ is a finite discrete group. The dual CFT resides on a
manifold with topology $R\times S^1\times {\cal M}_{p-1}$, where
$S^1$ represents the period of the coordinate $x$. Taking the limit
$\tilde{\alpha} \rightarrow 0$, we obtain Ricci flat AdS black holes
in general relativity
\begin{equation}
\label{2eq5} V_b(r) =\frac{r^2}{l^2} -\frac{16\pi G M_b}{p \eta_b
\Sigma r^{p-1}}.
\end{equation}
When $M_b = 0$, the solution reduces to
\begin{eqnarray}
\label{2eq6} && ds^2 = -V_r(r) dt^2 +V_r(r)^{-1} dr^2 +r^2 (dx^2
+h_{ij}dx^idx^j), \nonumber \\
&& V_r(r) = \frac{r^2}{2\tilde \alpha} \left
(1-\sqrt{1-\frac{4\tilde \alpha}{l^2}} \right ).
\end{eqnarray}
This is an AdS space in the Poincare coordinates with the effective
cosmological constant radius $l^2_{\rm eff}=2 \tilde \alpha
/(1-\sqrt{1-4\tilde\alpha/l^2})$. In addition, we see that the
Gauss-Bonnet coefficient must satisfy the condition $4\tilde
\alpha/l^2\ \le 1$, otherwise the theory is not well-defined.

 The black hole horizon $r_+$ is determined by
$V_b(r_+)=0$. Then the black hole mass can be expressed in terms of
the horizon $r_+$ as
\begin{equation}
\label{2eq7}
 M_b= \frac{p\eta_b \Sigma r_+^{p+1}}{16\pi G l^2}.
\end{equation}
The Hawking temperature of the black hole can be obtained by Wick
rotating the black hole solution (\ref{2eq3}) to its Euclidean
sector
\begin{equation}
\label{2eq8}
 ds^2= V_b(r) d\tau^2 +V_b(r)^{-1} dr^2 +r^2 (dx^2 +h_{ij}dx^idx^j).
 \end{equation}
 To remove the conical singularity at $r_+$ in the plane $(\tau,r)$, the
 Euclidean time $\tau$ must have a period $\beta_b$,
 \begin{equation}
 \label{2eq9}
 \beta_b \equiv 1/T_b= \frac{4\pi l^2}{(p+1) r_+},
 \end{equation}
 which is just the inverse Hawking temperature $1/T_b$ of the black
 hole. To calculate the Euclidean action of the black hole
 (\ref{2eq4}) and to regularize the action, we choose the vacuum
 solution (\ref{2eq6}) as a reference background.  In order that
 the Euclidean black hole solution (\ref{2eq8}) can be self-consistently embedded
 to the reference background, the Euclidean time of the vacuum
 solution must have a period $\beta_r$, which obeys
 \begin{equation}
 \beta_r \sqrt{V_r(r_b)}=\beta_b \sqrt{V_b(r_b)},
 \end{equation}
 where $r_b$ is the radial radius for a time-like
 hypersurface $(r=r_b>r_+)$, which acts as the boundary of the
 system. At the end of calculations, we will take the limit $r_b\rightarrow \infty$.
 Then the difference of the two Euclidean actions is~\footnote{In
 order to have a well-defined variable principle, there exist some surface terms in the action
 (\ref{2eq1}). As the case of general relativity, however, those
 surface terms have no contributions to the difference of Euclidean
 actions here.}
 \begin{eqnarray}
 \label{2eq11}
 {\cal I}_b &\equiv& I_b-I_r \nonumber \\
  &=& - \frac{\eta_b \Sigma \beta_b}{16\pi G} \int_{r_+}^{r_b} dr\ r^p\left( R +\frac{p
  (p+1) }{l^2} +\alpha R_{GB}\right) \nonumber \\
 && + \frac{\eta_b \Sigma \beta_r}{16\pi G} \int_0 ^{r_b} dr\ r^p\left( R
+\frac{p
  (p+1) }{l^2} +\alpha R_{GB}\right).
  \end{eqnarray}
Using (\ref{2eq2}), we have $\alpha R_{GB} =-(pR
+p(p+1)(p+2)/l^2)/(p-2)$. And notice that for the metric
(\ref{2eq3}), one has $R= -(r^pV_b)''/r^p$. Calculating
(\ref{2eq11}) and taking the limit $r_b\rightarrow \infty$, we
obtain
\begin{equation}
\label{2eq12}
 {\cal I}_b=-\frac{\eta_b \Sigma \beta_b}{16\pi
 G}\frac{r_+^{p+1}}{l^2} = -\frac{\eta_b \Sigma}{16\pi
 G l^2} \left(\frac{4 \pi l^2}{p+1} \right)^{p+1} \frac{1}{\beta_b^p}.
 \end{equation}
 The thermal energy can be calculated via the formula
 \begin{eqnarray}
 \label{2eq13}
 {\cal E}_b &\equiv& \frac{\partial {\cal I}_b}{\partial \beta_b}
   \nonumber \\
    &=& \frac{p\eta_b \Sigma }{16\pi G} \frac{r_+^{p+1}}{l^2}=M_b,
    \end{eqnarray}
which gives the black hole mass (\ref{2eq7}). The entropy of the
black hole can be obtained via $S=\beta_b {\cal E}_b-{\cal I}_b$,
and it gives
\begin{equation}
\label{2eq14}
 S=\frac{\eta_b\Sigma}{4 G} r_+^{p}=\frac{A}{4G},
\end{equation}
where $A$ is the horizon area. Very interestingly, although higher
order derivatives appear in the gravity action (\ref{2eq1}), the
entropy of the Ricci flat black holes still obeys the so-called area
formula. This is the feature of Ricci flat horizon. This feature
persists for Ricci flat black holes in more general Lovelock
gravity~\cite{Love}. For black hole horizons with positive or
negative constant curvature, the area formula does no longer hold in
Gauss-Bonnet gravity. In addition, let us calculate the heat
capacity of the black hole
 \begin{equation}
 \label{2eq15}
 C= \frac{\partial M_b}{\partial T_b}= \frac{p\eta_b
 \Sigma}{4G}r_+^p.
 \end{equation}
The heat capacity is always positive, which indicates that the black
hole can make thermal equilibrium with the surrounding thermal bath.
The negative definiteness of the Euclidean action (\ref{2eq12})
implies that the black hole is globally stable, and that the dual
CFT is in the deconfinement phase. Unlike its spherical black hole
counterpart, the Hawking-Page phase transition does not appear here.

In \cite{HM}, Horowitz and Myers found that there exists a
so-called AdS soliton solution in Einstein gravity with a negative
cosmological constant, which has a lower mass than the AdS vacuum.
The AdS soliton is obtained through a double-analytical
continuation. In our case, naturally we may expect that the AdS
vacuum (\ref{2eq6}) is not the lowest mass solution; following
\cite{HM} we can obtain the AdS soliton in the Gauss-Bonnet
gravity via analytically continuing the Ricci flat black hole
(\ref{2eq3}) with $t\rightarrow i \varphi$ and $x\rightarrow i t$.
Then we get a new solution
\begin{equation}
\label{2eq16} ds^2=V_s(r) d\varphi^2 +V_s(r)^{-1} dr^2 +r^2 (-dt^2
+h_{ij}dx^idx^j),
 \end{equation}
 where
 \begin{equation}
 \label{2eq17}
V_s(r)= \frac{r^2}{2\tilde\alpha}\left ( 1 -
 \sqrt{1+\frac{4\tilde \alpha r_s^{p+1}}{l^2 r^{p+1}}-
   \frac{4\tilde\alpha}{l^2}} \right)
   \end{equation}
   This solution is just the AdS soliton counterpart in Gauss-Bonnet
   gravity. Obviously there does not exist any horizon in the solution,
   but a conical singularity at $r=r_s$, which obeys
   $V_s(r_s)=0$.
 To remove this singularity, the coordinate $\varphi$ must have a
period $\beta_s$,
\begin{equation}
 \beta_s= \frac{4\pi l^2}{(p+1) r_s}.
 \end{equation}
 This solution is asymptotically AdS, and dual CFT now resides on
 manifold with topology $S^1\times R \times {\cal M}_{p-1}$, where
 $S^1$ denotes the period of the coordinate $\varphi$. In
 addition, let us mention here that the radial coordinate $r$ for
 the soliton solution ranges from $r_s$ to $\infty$.

 For the soliton solution (\ref{2eq17}), a natural reference
 background is the case with $r_s=0$, namely to replace $V_s$ in (\ref{2eq16}) by
 the same $V_r(r)$ given in (\ref{2eq6}). Again, to match the vacuum background,
 the period $\beta_r$ of the coordinate $\varphi$ for the reference
 background must obey the condition, $\beta_r \sqrt{V_r(r_b)} =\beta_s
 \sqrt{V_s(r_b)}$, on the boundary. With the solution $V_r$ as the
 reference background, and noticing that the Euclidean time for the AdS soliton  (\ref{2eq16})
 and the reference background can have
 an arbitrary period $\beta$,
  we find that the Euclidean action for the deformed AdS soliton solution (\ref{2eq16})
  is
  \begin{equation}
  {\cal I}_s= -\frac{\beta_s \Sigma \beta}{16\pi
  G}\frac{r_s^{p+1}}{l^2}.
  \end{equation}
  The corresponding mass is
 \begin{equation}
 \label{2eq20}
 {\cal E}_s  \equiv \frac{\partial {\cal I}_s}{\partial \beta}= -\frac{\beta_s \Sigma r_s^{p+1}}{16\pi G
 l^2},
\end{equation}
and as expected, the associated entropy vanishes, $S=\beta {\cal
E}_s-{\cal I}_s= 0$.

Note that the mass  (\ref{2eq20}) for the AdS soliton is indeed
negative, namely the mass of the soliton is less than the reference
background AdS space (\ref{2eq6}). On the other hand, let us notice
that for the same boundary topology $R\times S^1 \times {\cal
M}_{p-1}$, there exist three kinds of bulk solutions: Ricci flat
black hole solution (\ref{2eq3}), AdS space (\ref{2eq6}), and  the
deformed AdS soliton solution (\ref{2eq16}). From the point of view
of the dual CFTs, three kinds of bulk solutions might correspond to
three different phases, among which some phase transition may happen
like the Hawking-Page phase transition in the bulk for spherical AdS
black holes and confinement/deconfinement phase transition for the
dual CFTs. From the above calculations, however, we have seen that
unlike the spherical AdS black hole, there does not exist a phase
transition between the Ricci flat black hole and AdS space, and the
same is true in the case between the AdS soliton and AdS space.
Furthermore, let us point out that as in general relativity, the AdS
soliton has a less energy than the AdS space, therefore it is more
natural to consider the AdS soliton as the reference background. It
is quite interesting to see whether there does exist or not any
phase transition between the Ricci flat black hole and AdS soliton.
To see this, let us calculate the Euclidean action of the Ricci flat
black hole by viewing the deformed AdS soliton (\ref{2eq16}) as the
reference background. In order that the Ricci flat black hole can be
embedded into the AdS soliton background at the boundary $r=r_b$, in
the Euclidean sector, the Euclidean time period $\beta_{\tau}$ for
the AdS soliton and the period $\eta_b$ for the coordinate $x$ in
the Ricci flat black hole must obey the following conditions
\begin{equation}
\label{2eq21}
 r_b\beta_{\tau}= \beta_b \sqrt{V_b(r_b)}, \ \ \ r_b
\eta_b = \beta_s \sqrt{V_s(r_b)}.
\end{equation}
With the same procedure, we get the Euclidean action of the black
hole
 \begin{equation}
 \label{2eq22}
 {\cal I}_{bs} = -\frac{\Sigma \beta_b\beta_s}{16\pi Gl^2 l_{\rm
 eff}} \left( r_+^{p+1}-r_s^{p+1}\right).
 \end{equation}
 Here $l_{\rm eff}=\sqrt{2\tilde \alpha/ (1-\sqrt{1-4\tilde
 \alpha/l^2})}$, and its appearance is due to the fact that the
 coordinate $x$ for the Ricci flat black hole (\ref{2eq3}) and the  coordinate
 $\varphi$ in the deformed AdS
 soliton (\ref{2eq16}) have different dimensions. If we rescale
 $x$ in (\ref{2eq16}) as $x/l_{\rm eff}$, the factor $1/l_{\rm
 eff}$ in (\ref{2eq22}) will disappears.

 In this case, the mass of the black hole is
 \begin{equation}
 \label{2eq23}
 {\cal E}_{bs} \equiv \frac{\partial {\cal I}_{bs}}{\partial
 \beta_ b} = \frac{\Sigma\beta_s}{16\pi G l^2 l_{\rm eff}}
 \left(pr_+^{p+1}+r_s^{p+1}\right).
 \end{equation}
 And the associated entropy
 \begin{equation}
 S = \frac{\Sigma \beta_s}{4G l_{\rm eff}}r_+^p.
 \end{equation}
 Note that from (\ref{2eq21}) one has  $\eta_b = \beta_s/l_{\rm
 eff}$. Compared to (\ref{2eq13}) and (\ref{2eq14}), as expected,
 we see that the black hole mass depends on the choice of the
 reference background, but not for the entropy of the black hole.
 According to the Euclidean action (\ref{2eq22}), we see that when
$r_+>r_s$, it is negative, while it is positive as $r_s >r_+$.
That is, when crossing the boundary $r_+=r_s$, the Euclidean
action changes its sign.  The change of sign of the Euclidean
action is nothing but the indication of a first order phase
transition, as in the case of spherical AdS black hole.  But, there
is a significant difference between the cases of Ricci flat black
holes and spherical AdS black holes. For the Hawking-Page phase
transition of spherical AdS black holes, the phase transition is
determined by the ratio of black hole horizon $(r_+)$ and AdS
radius $l$: when $r_+>l$, the black hole phase dominates, while
the thermal AdS space dominates as $r_+<l$.  In our case, we can
see from (\ref{2eq22}) that the phase transition is now determined by
the ratio $r_+/r_s$.

In addition, recalling $\eta_b=\beta_s/l_{\rm eff}$, let us notice
that the Euclidean action (\ref{2eq22}) is the exactly the same as
in the case without the Gauss-Bonnet term~\cite{SSW}, although the
spacetime metrics are changed. As a result, the Gauss-Bonnet term
has no effect on the Hawking-Page phase transition. It is expected
that the result also holds for the Ricci flat black holes in the
more general Lovelock gravity.

\sect{Deformed AdS Soliton and Ricci flat black holes in dilaton
gravity}

In this section we will first consider a special kind of dilaton
gravity, which comes from spherical reduction of Dp-branes in type
II supergravity.  In this kind of dilaton gravity,  there exists a
kind of domain wall solutions, and the five-dimensional AdS space
appears as a special case. For this kind of domain wall
configurations, the holographic principle of quantum gravity can
be nicely illustrated via the so-called domain wall/QFT (quantum
field theory) correspondence~\cite{DW}, which generalizes the
AdS/CFT correspondence to the case of non-conformal quantum
theories.

Let us start from the action of type II supergravity in the string
frame
\begin{equation}
\label{3eq1}
 S=\frac{1}{2\pi G_{10}}\int d^{10}x\sqrt{-g} \left( e^{-2\phi}(R
 +4(\partial \phi)^2) - \frac{1}{2(8-p)!}F^2_{p-2}\right),
 \end{equation}
 where $G_{10}= 8\pi^6 \alpha'^4$ is the gravitational constant in
 ten dimensions. The black Dp-brane solution in the string frame
 has the form
 \begin{eqnarray}
 \label{3eq2}
 && ds^2_{\rm string}=H^{-1/2}(-f dt^2 +dx_p^2)
 +H^{1/2}(f^{-1}dr^2 +r^2 d\Omega_{8-p}^2),
   \nonumber \\
 && e^{\phi} =g_s H^{(3-p)/4}, \nonumber \\
&& F_{8-p}= Q \epsilon_{8-p},
\end{eqnarray}
where $0 \le p \le 6$, $g_s$ is the string coupling constant at
infinity, $\epsilon_{8-p}$ is the volume form of the sphere
$S^{8-p}$, and $Q$ is the magnetic charge of the Dp-brane. In
addition,
\begin{equation}
H= 1+\frac{r_0^{7-p}\sinh^2 \alpha}{r^{7-p}}, \ \ \ f=
1-\frac{r_0^{7-p}}{r^{7-p}}.
\end{equation}
In the decoupling limit, $\alpha' \rightarrow 0$, but keeping fixed
$U=r/\alpha'$, $U_0=r_0/\alpha'$ and the Yang-Mills coupling
constant $g^2_{\rm YM}$, with $g^2_{\rm YM}= g_s
(\alpha')^{(p-3)/2}$, the harmonic function reduces to
\begin{equation}
H= \frac{g^2_{\rm YM} N}{\alpha'^2 U^{7-p}},
\end{equation}
where $N$ is the number of Dp-branes and we have absorbed a
numerical coefficient into the Yang-Mills coupling constant
$g^2_{\rm YM}$. Except for the case of $p=3$, the radius of angular
part in the string metric (\ref{3eq2}) depends on $U$. In order to
remove the dependence, let us consider the so-called ``dual
frame"~\cite{DW}
\begin{equation}
ds^2_{\rm dual}= (Ne^{\phi})^{2/(p-7)} ds^2_{\rm string}.
\end{equation}
In this frame, the action turns out to be~\cite{CO}
 \begin{equation}
 \label{3eq6}
 S= \frac{N^2}{16\pi G_{10}} \int d^{10}x\sqrt{-g}
 (Ne^{\phi})^{\lambda} \left (R
 +\frac{4(p-1)(p-4)}{(7-p)^2}(\partial
 \phi)^2-\frac{1}{2N^2(8-p)!}F^2_{8-p}\right),
 \end{equation}
 where $\lambda= 2(p-3)/(7-p)$. In the decoupling limit, the
 solution in the dual frame has the form
 \begin{eqnarray}
 \label{3eq7}
 && ds^2_{\rm dual} = \alpha' \left( (g^2_{\rm YM}N)^{-1}
 U^{5-p}(-fdt^2 +dx_p^2) +U^{-2}f^{-1} dU^2 +d\Omega_{8-p}^2
 \right), \nonumber \\
 && e^{\phi}= \frac{1}{N} (g^2_{\rm YM}N U^{p-3})^{(7-p)/4},
 \nonumber \\
 && F_{8-p}= (7-p) N (\alpha')^{(7-p)/2}\epsilon_{8-p},
 \end{eqnarray}
 where $f=1-(U_0/U)^{7-p}$. The metric is of the form
 $AdS_{p+2}\times S^{8-p}$ for $p\ne 5$ and $E^{(1,6)} \times S^3$
 for $p=5$ in the dual frame.  Note that for the case of $p=3$,
 these two frames are equivalent to each other since the dilaton
 is a constant in this case. In addition, let us notice that in
 the dual frame, the radius of the angular part of the metric is a
 constant. Therefore in this frame we can conveniently do a
 spherical reduction of the type II supergravity on $S^{8-p}$, and
 obtain the effective action of the gauged supergravity in the
 Einstein frame~\cite{CO}~\footnote{since $\alpha'$ will be eventually cancelled
 at the end of calculations, we will set $\alpha'=1$ in the
 following.}
 \begin{equation}
 \label{3eq8}
 S= \frac{N^2\Omega_{8-p}}{(2\pi)^7}\int d^{p+2}x\sqrt{-g} ( R
   -\frac{1}{2} (\partial \Phi)^2
   +V(\Phi))-\frac{2N^2\Omega_{8-p}}{(2\pi)^7}\int
   d^{p+1}x\sqrt{-h}K,
   \end{equation}
   where we have added a Gibbons-Hawking surface term,
   $\Omega_{8-p}$ is the volume of the unit sphere $S^{8-p}$, and
   \begin{eqnarray}
   \label{3eq9}
   && V(\Phi) =\frac{1}{2} (9-p)(7-p) N^{-2\lambda/p}e^{a\Phi},
   \nonumber \\
   && \Phi= \frac{2\sqrt{2(9-p)}}{\sqrt{p}(7-p)}\phi, \ \ \
   a=-\frac{\sqrt{2}(p-3)}{\sqrt{p(9-p)}}.
   \end{eqnarray}
   After the reduction, we obtain a Ricci flat black hole (black
   domain wall) solution
   \begin{eqnarray}
   \label{3eq10}
   && ds^2= (Ne^{\phi})^{2\lambda/p}\left [(g^2_{\rm YM}N)^{-1}
 U^{5-p}(-fdt^2 +dx_p^2) +U^{-2}f^{-1} dU^2
 \right ], \nonumber \\
 && e^{\phi}= \frac{1}{N} (g^2_{\rm YM} U^{p-3})^{(7-p)/4}.
 \end{eqnarray}
 The Ricci flat black hole has a horizon at $U=U_0$, where $f(U)$ vanishes.
 When $p\ne 5$, we can make a further transformation
 \begin{equation}
 \label{3eq11}
 u^2 = {\cal R}^2 (g^2_{\rm YM}N)^{-1} U^{5-p}, \ \ {\cal R}
 =2/(5-p),
 \end{equation}
 so that the Ricci flat solution takes the form
 \begin{eqnarray}
 \label{3eq12}
 && ds^2 = (Ne^{\phi})^{2\lambda/p}\left[ \frac{u^2}{{\cal R}^2}
 \left(-\tilde f dt^2+dx_p^2 \right) +\frac{{\cal R}^2}{u^2 \tilde
 f}du^2 \right], \nonumber\\
 && e^{\phi}= \frac{1}{N} (g^2_{\rm
 YM}N)^{(7-p)/2(5-p)}(\frac{u}{\cal R})^{(p-7)(p-3)/2(p-5)},
 \nonumber\\
 && \tilde f = 1-\left(\frac{u_0}{u}\right)^{2(7-p)/(5-p)},
 \end{eqnarray}
 where $u^2_0= {\cal R}^2 (g^2_{\rm YM}N)^{-1} U_0^{5-p}$. Clearly
 the Ricci flat black hole is conformal to $AdS_{p+2}$.

In \cite{CO}, the stress-energy tensor of dual quantum field theory
and the mass of the Ricci flat black hole have been calculated via
the surface counterterm method. The counterterm is found to be
\begin{equation}
\label{3eq13}
 S_{\rm ct} =-\frac{2N^2 \Omega_{8-p}}{(2\pi)^7} \int
d^{p+1}\sqrt{-h} \frac{c_0}{l_{\rm eff}},
\end{equation}
where
\begin{equation}
c_0 =\sqrt{\frac{(9-p)p(p+1)}{2(7-p)}}, \ \ \ \frac{1}{l_{\rm
eff}} = \sqrt{\frac{V(\Phi)}{p(p+1)}}.
 \end{equation}
 The mass of the black hole is
 \begin{equation}
 \label{3eq15}
 M = \frac{\Omega_{8-p}}{(2\pi)^7g^4_{\rm YM}}
 \frac{9-p}{2} U_0^{7-p}V_p,
 \end{equation}
 where $V_p$ is the volume for the Euclidean space described by
 $dx_p^2$. The Euclidean action of the Ricci flat black hole is
 \begin{eqnarray}
 \label{3eq16}
 {\cal I} &=& -\frac{N^2\Omega_{8-p}}{(2\pi)^7} \int
 d^{p+2}x\sqrt{g} \left (R -\frac{1}{2} (\partial \Phi)^2 +V(\Phi)
 \right) \nonumber \\
  && + \frac{2N^2\Omega_{8-p}}{(2\pi)^7} \int d^{p+1}x \sqrt{h} K
  +\frac{2N^2\Omega_{8-p}}{(2\pi)^7}\int d^{p+1}x \sqrt{h}
  \frac{c_0}{L_{\rm eff}}, \nonumber \\
  &=& -\frac{\Omega_{8-p}V_p U_0^{7-p}}{(2\pi)^7 g^4_{\rm YM}T}
  \frac{5-p}{2},
  \end{eqnarray}
where $T$ is the Hawking temperature of the black hole
\begin{equation}
\label{3eq17}
 T = \frac{7-p}{4\pi } \frac{1}{\sqrt{g^2_{\rm
YM}N}}U_0^{\frac{5-p}{2}}.
\end{equation}
According to ${\cal E}= \partial {\cal I}/\partial \beta$ and
$S=\beta {\cal E}-{\cal I}$, it is easy to see that the energy of
the black hole is ${\cal E}=M$ and the entropy of the black hole
\begin{equation}
\label{3eq18}
 S= \frac{\Omega_{8-p}V_p}{2^5\pi^6 g^4_{\rm
YM}}\sqrt{g^2_{\rm YM}N} U_0^{(9-p)/2}.
\end{equation}
It is easy to show that the surface counterterm method here is
equivalent to the background subtraction method if one chooses the
solution (\ref{3eq10}) with $U_0=0$ or (\ref{3eq12}) with $u_0=0$ as
the reference background. We see from (\ref{3eq16}) that the action
is always negative for $0\le p<5$,  positive for $p>5$, and vanishes
for the case of $p=5$.  This implies that compared to the thermal
dilaton background (\ref{3eq10}) with $U_0=0$ or (\ref{3eq12}) with
$u_0=0$,  the Ricci flat black hole is globally stable and dominates
when $p<5$, is marginal when $p=5$, and unstable for the case $p>5$.
Notice the fact that the supergravity configuration is dual to the
little string theory for the case $p=5$, while gravity does not be
decoupled for the case $p>5$. Therefore our results are consistent
with that observation. Furthermore, we see from the Euclidean action
(\ref{3eq16}) that there does not exist any phase transition between
the Ricci flat black holes and thermal dilaton background.

Now we consider double-analytically continuing the Ricci flat
black holes with $t\rightarrow i\varphi$ and $x_1 \rightarrow it$,
where $x_1$ is one of coordinates $x_p$. \footnote{In that case, one
has to have $p\ge 1$, namely we exclude the black D0-brane
solution here.} In that case, the  black hole solution becomes
\begin{equation}
\label{3eq19} ds^2_s= (Ne^{\phi})^{2\lambda/p}\left [(g^2_{\rm
YM}N)^{-1}
 U^{5-p}(fd\varphi^2 -dt^2 +dx_{p-1}^2) +U^{-2}f^{-1} dU^2
 \right ]
 \end{equation}
 or
 \begin{equation}
 \label{3eq20}
 ds^2_s = (Ne^{\phi})^{2\lambda/p}\left[ \frac{u^2}{{\cal R}^2}
 \left(\tilde f d\varphi^2-dt^2 +dx_{p-1}^2 \right) +\frac{{\cal R}^2}{u^2 \tilde
 f}du^2 \right],
 \end{equation}
 where $U_0$ ($u_0$) is replaced by $U_s(u_s)$ and the dilaton field is still given by (\ref{3eq10}) or
 (\ref{3eq12}). For this solution, the horizon disappears and in
 order to remove the conical singularity, the coordinate $\varphi$
 has to have a period with
 \begin{equation}
 \label{3eq21}
 \beta_s=\frac{4\pi}{7-p}\sqrt{g^2_{\rm YM}N }
 U_s^{\frac{p-5}{2}}.
 \end{equation}
 In addition, let us notice that to keep the signature of the solution,
 one has to take $U\ge U_s$ in (\ref{3eq19}) or $u\ge u_s$ in (\ref{3eq20}).
 Note also that the solution (\ref{3eq20}) is just the AdS soliton when
 $p=3$. Naturally, we call the solution (\ref{3eq19}) or
 (\ref{3eq20}) the dilaton deformed AdS soliton. As in the
 case of the AdS soliton, we expect that the dilaton deformed AdS
 soliton has a lower mass than the dilaton background (\ref{3eq10})
 with
 $f=\tilde f=1$. To see this, let us calculate the Euclidean
 action of the Ricci flat black hole by viewing the soliton
 solution as the reference background, which gives us with
 \begin{equation}
 \label{3eq22}
 {\cal I}_{bs}= -\frac{\Omega_{8-p}}{(2\pi)^7}\frac{\beta_s\beta
 V_{p-1}}{g^4_{\rm YM}}\frac{5-p}{2} (U_0^{7-p}-U_s^{7-p}),
 \end{equation}
 where $\beta=1/T$ is the inverse Hawking temperature (\ref{3eq17})
  of the Ricci flat black hole. From the Euclidean action, we can
  clearly see that there does exist the Hawking-Page phase
  transition between the Ricci flat black hole and dilaton
  deformed AdS soliton. When $1 \le p<5$, the black hole phase is
  globally stable and dominates when $U_0>U_s$.  When $p=6$,
  however, the thermal soliton phase is globally stable and
  dominates when $U_s>U_0$. This difference is caused by the
  relation between the Hawking temperature and $U_0$
  (\ref{3eq17}): when $p<5$,  a larger black hole has a higher
  temperature, while the converse happens for the case of $p=6$.
  When $p=5$, the situation is subtle, the temperature of the
  black hole is a constant, and the Euclidean action always vanishes.
  From the point of view of dual little string theory, one has to
  consider higher order corrections~\cite{Harm}. Viewing the deformed soliton
  as the reference background, the mass of the black hole is
  \begin{equation}
  \label{3eq23}
  {\cal
  E}_{bs}=\frac{\Omega_{8-p}}{2(2\pi)^7}\frac{\beta_sV_{p-1}}
  {g^4_{\rm YM}}\left ((9-p)U_0^{7-p}+(5-p) U_s^{7-p}\right),
\end{equation}
and associated entropy with the black hole is still given by
(\ref{3eq18}).

So far we have considered the dilaton gravity (\ref{3eq8}) with
the coupling constant $a$ given by (\ref{3eq9}), which comes from
the spherical reduction of Dp-brane in type II supergravity.  Next
we consider a dilaton gravity with a Liouville-type potential with
an arbitrary coupling constant $a$,
\begin{equation}
\label{3eq24}
 S = \frac{1}{16\pi G} \int d^{p+2}x\sqrt{-g}
\left(R-\frac{1}{2} (\partial \phi)^2 +V_0 e^{-a\phi}\right),
\end{equation}
where $G$ is the gravitational constant in $p+2$ dimensions and
$V_0$ is a constant. The Ricci flat black hole solution has the
form~\cite{CZ}
\begin{eqnarray}
\label{3eq25}
&& ds^2 = -f(r)dt^2 +f(r)^{-1} dr^2 +R^2(r)dx_p^2, \nonumber\\
&& R(r)=r^n, \nonumber \\
&& \phi(r) =\phi_0 +\sqrt{2np(1-n)}\ln r, \nonumber \\
&& f(r) = \frac{V_0 e^{-a\phi_0}r^{2n}}{ np(n(p+2)-1)}- {\rm \bf
m}_b r^{1-np},
\end{eqnarray}
where $\phi_0$ and ${\rm \bf m}_b$ are two integration constants,
$dx_p$ describes the line element of $p$-dimensional Ricci flat
space, and the constant $n$ has a relation to the coupling
constant $a$ as
\begin{equation}
a = \frac{\sqrt{2np(1-n)}}{np}.
\end{equation}
Note that when $n=1$, the solution (\ref{3eq25}) reduces to the
$(p+2)$-dimensional AdS Ricci flat black hole with a constant
dilaton. Therefore we will not consider the special case with
$n=1$ in what follows. However, the case of $n=1$ will be
naturally included in the following discussions. The black hole
horizon $r_+$ is determined by the equation $f(r)|_{r=r_+}=0$. The
associated Hawking temperature is
\begin{equation}
\label{3eq27} T=1/\beta= \frac{V_0e^{-a\phi_0}}{4\pi np}
r_+^{2n-1}.
\end{equation}
The entropy of the black hole obeys the so-called area formula
since we are working in the Einstein frame, and is
\begin{equation}
\label{3eq28}
 S=\frac{V_p}{4G}r_+^{np},
 \end{equation}
 where $V_P$ is the volume of the manifold $dx^2_p$. Furthermore,
 if we choose the solution with ${\rm \bf m}_b=0$ as the
reference background, the mass of the black hole has the form
\begin{equation}
\label{3eq29}
 M=\frac{{\rm \bf m}_b V_p np}{16\pi G}=
 \frac{V_0e^{-a\phi_0} V_p r_+^{n(p+2)-1}}{16\pi G (n(p+2)-1)}.
 \end{equation}
 The Euclidean action of the black hole can also be calculated
 by the formula: ${\cal I}  \equiv \beta {\cal F}= \beta M-S$,
 where ${\cal F}$ is the free energy of the black hole, which is
 equivalent to obtain the Euclidean action from (\ref{3eq24}). We
 find
 \begin{equation}
 \label{3eq30}
 {\cal I} =-\frac{V_p r_+^{np}}{4G} \frac{2n-1}{n(p+2)-1}.
 \end{equation}
 The action is always negative if $n>1/2$ or $n<1/(p+2)$, and
 positive if $ 1/(p+2) <n <1/2$. Let us notice that in order to have
 a well-behaved boundary metric, on which the dual QFT resides, we have
 $n>1/(p+2)$ from the solution (\ref{3eq25}). In addition, when $n=1/2$, the action
 vanishes like the case of D5-branes.

 The dilaton deformed AdS soliton solution for the action
 (\ref{3eq24}) can be obtained by double analytical continuation
 from the black hole solution (\ref{3eq25}) via $t \rightarrow
 i\varphi$ and $x_1 \rightarrow it$, so that we have
 \begin{eqnarray}
 \label{3eq31}
&& ds^2 = f(r)d\varphi ^2 +f(r)^{-1} dr^2 +R^2(r)(-dt^2 +dx_{p-1}^2), \nonumber\\
&& R(r)=r^n, \nonumber \\
&& \phi(r) =\phi_0 +\sqrt{2np(1-n)}\ln r, \nonumber \\
&& f(r) = \frac{V_0 e^{-a\phi_0}r^{2n}}{ np(n(p+2)-1)}- {\rm \bf
m}_s r^{1-np},
\end{eqnarray}
To remove the conical singularity at $r=r_s$, which satisfies
$f(r_s)=0$, the coordinate $\varphi$ has to have a period
$\beta_s$ obeying
\begin{equation}
\label{3eq32}
 \beta_s =\frac{4\pi np}{V_0e^{-a\phi_0}}r_s^{1-2n}.
 \end{equation}
 Viewing the soliton as the reference background, we obtain the
 Euclidean action of the black hole
 \begin{equation}
 \label{3eq33}
 {\cal I}_{bs}= -\frac{V_0e^{-a\phi_0}V_{p-1}\eta_b \beta}{16\pi
 G\
 np\ }\frac{2n-1}{n(p+2)-1}\left(r_+^{n(p+2)-1}-r_s^{n(p+2)-1}\right),
 \end{equation}
 where $\eta_b$ is the period of the coordinate $x_1$ for the
 black hole solution (\ref{3eq25}), which has a relation to
 $\beta_s$ via $\eta_b=\beta_s/l_{\rm eff}$ with
 $l_{\rm eff}= \sqrt{\frac{np(n(p+2)-1)}{V_0e^{-a\phi_0}}}$.
 Again, in this background, the energy of the black hole is
 \begin{equation}
 \label{3eq34}
 {\cal E}_{bs}=\frac{V_0e^{-a\phi_0}\eta_bV_{p-1}}{16\pi G
 np}\frac{1}{n(p+2)-1}\left( np \ r_+^{n(p+2)-1}+(2n-1)
 r_s^{n(p+2)-1}\right).
 \end{equation}
 And the entropy is still given by (\ref{3eq28}). We see from the
 action (\ref{3eq33}) that there does exist a phase transition
 between the Ricci flat black hole and the dilaton deformed AdS
 soliton when $n\ne 1/2$. When $n>1/2$, the black hole phase
 dominates if $r_+>r_s$. When $n<1/2$, the thermal soliton phase
 dominates if $r_s >r_+$. As the Euclidean action
 changes its sign, a Hawking-Page phase transition happens.

\sect{Conclusions and Discussions}
In this paper we have studied Ricci flat black holes and deformed
 AdS soliton in Gauss-Bonnet gravity and dilaton gravity with a
 Louville-type dilaton potential including the gauged supergravity
 coming from the spherical reduction of Dp-branes in type II
 supergravity.  In Gauss-Bonnet gravity, the black hole
 solution and AdS soliton are greatly deformed by the Gauss-Bonnet
 term, but the black hole entropy still obeys the area formula and
 the Hawking-Page phase transition between the black hole and
 soliton background is still determined by the black hole
 temperature through the horizon $r_+$ and the compact radius
 $\eta_b$ through $r_s$. The Gauss-Bonnet coefficient $\alpha$
 explicitly disappears in the Euclidean action. As a result, the
 Gauss-Bonnet term has no effect on the Hawking-Page phase
 transition, compared to the case without the Gauss-Bonnet term.
 This is quite different from the case of spherical black
 holes~\cite{GB}.

 In dilaton gravity, the high temperature phase is dominated by
 black holes in some cases, and in other cases is dominated by deformed
 thermal solitons,  depending on the dilaton coupling constant.
 For example, see the action (\ref{3eq22}). When $p<5$, the high
 temperature phase is dominated by the black hole while by the
 soliton background for $p=6$. The same happens in (\ref{3eq33}):
 in the high temperature phase, the black hole dominates when
 $n>1/2$, while the deformed AdS soliton dominates when $n<1/2$.
 This feature is new, compared to the case of Einstein gravity,
 where black hole always dominates in the high temperature phase~\cite{SSW}.

\section*{Acknowledgments}
The work of R.G.C. was supported in part by a grant from Chinese
Academy of Sciences, and by NSFC under grants No. 10325525 and No.
90403029. The work of S.P.K. was supported by the Korea Science and
Engineering Foundation (KOSEF) grant funded by the Korea government
(MOST) (No. R01-2005-000-10404-0). B.W. was  partially supported by
the grants from NSFC.


\begin{thebibliography}{99}

\bibitem{AdS}J.~Maldacena,
Adv.\ Theor.\ Math.\ Phys.\  {\bf 2}, 231 (1998) [Int.\ J.\
Theor.\ Phys.\  {\bf 38}, 1113 (1998)] [hep-th/9711200];

 S.~S.~Gubser, I.~R.~Klebanov and A.~M.~Polyakov,
Phys.\ Lett.\ B {\bf 428}, 105 (1998) [hep-th/9802109];

 E.~Witten,
Adv.\ Theor.\ Math.\ Phys.\  {\bf 2}, 253 (1998) [hep-th/9802150].

\bibitem{Witten} E.~Witten,
Adv.\ Theor.\ Math.\ Phys.\  {\bf 2}, 505 (1998) [hep-th/9803131].

\bibitem{HP}S.~W.~Hawking and D.~N.~Page,
Commun.\ Math.\ Phys.\  {\bf 87}, 577 (1983).


\bibitem{Topo}J.~L.~Friedman, K.~Schleich and D.~M.~Witt,
Phys.\ Rev.\ Lett.\  {\bf 71}, 1486 (1993) [Erratum-ibid.\  {\bf
75}, 1872 (1993)] [gr-qc/9305017];

T.~Jacobson and S.~Venkataramani,
Class.\ Quant.\ Grav.\  {\bf 12}, 1055 (1995)
[arXiv:gr-qc/9410023];

G.~J.~Galloway, K.~Schleich, D.~M.~Witt and E.~Woolgar,
Phys.\ Rev.\ D {\bf 60}, 104039 (1999) [arXiv:gr-qc/9902061];

E.~Woolgar,
Class.\ Quant.\ Grav.\  {\bf 16}, 3005 (1999)
[arXiv:gr-qc/9906096].

\bibitem{Flat}J.~P.~Lemos,
Class.\ Quant.\ Grav.\  {\bf 12}, 1081 (1995) [gr-qc/9407024];

J.~P.~Lemos,
Phys.\ Lett.\ B {\bf 353} (1995) 46 [gr-qc/9404041];

C.~Huang and C.~Liang,
Phys.\ Lett.\ A {\bf 201} (1995) 27;

C.G. Huang, Acta Phys. Sin. {\bf 4}, 617 (1995).

J.~P.~Lemos and V.~T.~Zanchin,
Phys.\ Rev.\ D {\bf 54}, 3840 (1996) [hep-th/9511188].

 R.~G.~Cai and Y.~Z.~Zhang,
Phys.\ Rev.\ D {\bf 54}, 4891 (1996) [gr-qc/9609065];

\bibitem{Negative}R.~B.~Mann,
Class.\ Quant.\ Grav.\  {\bf 14}, L109 (1997) [gr-qc/9607071];

W.~L.~Smith and R.~B.~Mann,
Phys.\ Rev.\ D {\bf 56}, 4942 (1997) [gr-qc/9703007];

R.~B.~Mann,
Nucl.\ Phys.\ B {\bf 516}, 357 (1998) [hep-th/9705223].


D.~R.~Brill, J.~Louko and P.~Peldan,
Phys.\ Rev.\ D {\bf 56}, 3600 (1997) [gr-qc/9705012].

L.~Vanzo,
Phys.\ Rev.\ D {\bf 56}, 6475 (1997) [gr-qc/9705004].

\bibitem{Birm}D.~Birmingham,
Class.\ Quant.\ Grav.\  {\bf 16}, 1197 (1999) [hep-th/9808032].

\bibitem{high}R.~Cai and K.~Soh,
Phys.\ Rev.\ D {\bf 59}, 044013 (1999) [gr-qc/9808067].

A.~Chamblin, R.~Emparan, C.~V.~Johnson and R.~C.~Myers,
Phys.\ Rev.\ D {\bf 60}, 064018 (1999) [hep-th/9902170].

R.~Aros, R.~Troncoso and J.~Zanelli,
Phys.\ Rev.\ D {\bf 63}, 084015 (2001) [hep-th/0011097].

S.~Nojiri and S.~D.~Odintsov,
Phys.\ Lett.\ B {\bf 521}, 87 (2001) [arXiv:hep-th/0109122];

M.~Cvetic, S.~Nojiri and S.~D.~Odintsov,
  Nucl.\ Phys.\  B {\bf 628}, 295 (2002)
  [arXiv:hep-th/0112045].


Y.~M.~Cho and I.~P.~Neupane,
  Phys.\ Rev.\  D {\bf 66}, 024044 (2002)
  [arXiv:hep-th/0202140];

 I.~P.~Neupane,
  Phys.\ Rev.\  D {\bf 67}, 061501 (2003)
  [arXiv:hep-th/0212092].



\bibitem{GB} R.~G.~Cai,
  Phys.\ Rev.\  D {\bf 65}, 084014 (2002)
  [arXiv:hep-th/0109133].

\bibitem{Love}R.~G.~Cai,
  Phys.\ Lett.\  B {\bf 582}, 237 (2004)
  [arXiv:hep-th/0311240].

\bibitem{CJS}R.~G.~Cai, J.~Y.~Ji and K.~S.~Soh,
  Phys.\ Rev.\  D {\bf 57}, 6547 (1998)
  [arXiv:gr-qc/9708063].

\bibitem{CZ}R.~G.~Cai and Y.~Z.~Zhang,
  Phys.\ Rev.\  D {\bf 64}, 104015 (2001)
  [arXiv:hep-th/0105214].

\bibitem{HM} G.~T.~Horowitz and R.~C.~Myers,
  Phys.\ Rev.\  D {\bf 59}, 026005 (1999)
  [arXiv:hep-th/9808079].

  \bibitem{SSW}S.~Surya, K.~Schleich and D.~M.~Witt,
  Phys.\ Rev.\ Lett.\  {\bf 86}, 5231 (2001)
  [arXiv:hep-th/0101134].

  \bibitem{Page} D.~N.~Page,
  arXiv:hep-th/0205001.

\bibitem{Myers}R.~C.~Myers,
  Phys.\ Rev.\  D {\bf 60}, 046002 (1999)
  [arXiv:hep-th/9903203].

\bibitem{Shen}J.~Shen, B.~Wang, C.~Y.~Lin, R.~G.~Cai and R.~K.~Su,
  arXiv:hep-th/0703102.

\bibitem{DW}H.~J.~Boonstra, K.~Skenderis and P.~K.~Townsend,
  JHEP {\bf 9901}, 003 (1999)
  [arXiv:hep-th/9807137];

  K.~Behrndt, E.~Bergshoeff, R.~Halbersma and J.~P.~van der Schaar,
  Class.\ Quant.\ Grav.\  {\bf 16}, 3517 (1999)
  [arXiv:hep-th/9907006].




\bibitem{CO}R.~G.~Cai and N.~Ohta,
  Phys.\ Rev.\  D {\bf 62}, 024006 (2000)
  [arXiv:hep-th/9912013].

\bibitem{Harm}T.~Harmark and N.~A.~Obers,
  Phys.\ Lett.\  B {\bf 485}, 285 (2000)
  [arXiv:hep-th/0005021].


\end{thebibliography}
\end{document}